\documentclass[12pt]{iopart}

\usepackage{graphicx, amssymb}

\begin{document}
\title{Stability window and mass-radius relation for magnetized strange quark stars}

\author{R. Gonz\'{a}lez Felipe$^{1,2}$, A. P\'{e}rez Mart\'{\i}nez$^{3,4}$}

\address{$^{1}$Instituto Superior de Engenharia de Lisboa\\
Rua Conselheiro Em\'{\i}dio Navarro, 1959-007 Lisboa, Portugal}
\address{$^{2}$Centro de F\'{\i}sica Te\'{o}rica de Part\'{\i}culas, Instituto Superior T\'{e}cnico\\
Avenida Rovisco Pais, 1049-001 Lisboa, Portugal}

\address{$^{3}$Centro Brasilero de Pesquisas Fisicas\\
Instituto de Cosmologia, Relatividade e Astrofisica (ICRA-BR),\\
Rua Dr. Xavier Sigaud 150, 9 Cep 22290-180, Urca, Rio de Janeiro, RJ, Brazil}
\address{$^{4}$Instituto de Cibern\'{e}tica, Matem\'{a}tica y F\'{\i}sica (ICIMAF) \\Calle E esq 15 No. 309 Vedado, Havana, 10400, Cuba}

\ead{gonzalez@cftp.ist.utl.pt, aurora@icmf.inf.cu}

\begin{abstract}
The stability of magnetized strange quark matter (MSQM) is investigated within the phenomenological MIT bag model, taking into account the variation of the relevant input parameters, namely, the strange quark mass, baryon density, magnetic field and bag parameter. We obtain that the energy per baryon decreases as the magnetic field increases, and its minimum value at vanishing pressure is lower than the value found for SQM. This implies that MSQM is more stable than non-magnetized SQM. Furthermore, the stability window of MSQM is found to be wider than the corresponding one of SQM. The mass-radius relation for magnetized strange quark stars is also derived in this framework.
\end{abstract}

\maketitle

\section{Introduction}
\label{sec1}

Strange quark matter (SQM), in contrast to ordinary nuclei (where quarks are confined into colorless nucleons), is thought to be a form of macroscopic matter which contains a large quantity of deconfined quarks ($u$, $d$ and $s$) in $\beta$-equilibrium, with electric and color charge neutrality. Theoretically, Bodmer~\cite{Bodmer:1971we} and Witten~\cite{Witten:1984rs} proved that SQM could be the more stable phase of nuclear matter. The reason behind this crucial conclusion is the presence of the quark $s$ with a mass lower than the Fermi energy, which leads to a decreasing of the total energy per baryon $E/A$, if some of the $u$ and $d$ quarks on the Fermi surface are converted into $s$ quarks via weak interactions. This is, in other words, a consequence of the Pauli exclusion principle.

If the hypothesis of the existence of SQM is true, this form of matter could have interesting astrophysical consequences. For instance, SQM could be realized in the inner core of compact objects, such as neutron stars, or even form the hypothetical strange stars. It may be possible as well that small quarks droplets could be formed in relativistic heavy ion collider (RHIC) experiments~\cite{RHIC}. As SQM is more stable and bound at finite density, one expects configurations of quark stars with macroscopic properties quite different from neutron stars~\cite{itoh,Hansel}. These objects would be self-bound and their masses would scale with the radius as $M \sim R^{3}$, in contrast to neutron stars which have masses that decrease with increasing radius ($M \sim R^{-3}$) and are bound by gravity. Thus, they could explain a set of peculiar astrophysical observations which are awaiting for theoretical explanations~\cite{Lugones:2002va, Lugones:2002zd,Quan,Nice,Xu}.

SQM has been widely studied using the MIT bag model, where the non-interacting quark gas at zero temperature is considered to be confined into a phenomenological bag, $B_{\rm bag}$, which mimics the strong interaction. Such studies have revealed that even heavy strangelets might be stable, with the corresponding $E/A$ lower than that of the iron nuclei, $E/A\,(^{56}{\rm Fe}) \simeq 930$~MeV~\cite{Farhi:1984qu}. Being a phenomenological model, the MIT bag model has some limitations that are well known. The confinement is put by hand, and the model is unable to reproduce the chiral symmetry breaking at zero density. Nevertheless, the model is quite satisfactory in describing quark matter at finite density.

Focusing on a real astrophysical scenario, one should also note the important role played by the strong magnetic fields expected in this context. Pulsars, magnetars, neutron stars, the emission of intense sources of X-rays could be associated to sources with intense magnetic fields around $10^{13}-10^{15}$~G or even higher fields~\cite{duncan,kouve}. We remark that the magnetic field intensity may vary significantly from the surface to the center of the source. Although there is no direct observational evidence for the magnetic field strength in the inner core of stars, theoretical estimates indicate that fields as high as $10^{18}$~G could be allowed. Therefore, SQM in the presence of strong magnetic fields is a subject that it is worth addressing. In a recent work~\cite{Felipe:2007vb}, we investigated the SQM properties in the presence of a strong magnetic field, taking into account $\beta$- equilibrium and the anomalous magnetic moments (AMM) of the quarks. It was found that the stability of the system requires a magnetic field value $B \lesssim 10^{18}$~G. This is in contrast to the bound $B \lesssim 10^{19}$~G, obtained when AMM are not included~\cite{Chakrabarty:1996te}.

The scope of the present paper is to establish a bridge between the microscopic properties of magnetized SQM (MSQM), given by the equation of state (EoS) of the system, and the macroscopic observables of strange quark stars (SQS), derived from this kind of magnetized matter. In the framework of the MIT bag model, we study the behavior of the system with the variation of its parameters, namely, the bag parameter $B_{\rm bag}$, the strange quark mass $m_s$, the baryon density $n_B$ and the magnetic field $B$. The possible mass-radius configurations of magnetized strange quark stars (MSQS) are then obtained by solving the Tolman-Oppenheimer-Volkoff (TOV) equation. As it turns out, stable stars with smaller radii are allowed due to the compactness of matter and the presence of a strong magnetic field, since the energy per baryon at vanishing pressure is lower in this case.

The paper is organized as follows. In section \ref{sec2} we briefly review the thermodynamic properties of SQM in the presence of a strong magnetic field. In section \ref{sec3} the stability windows of SQM in the presence of a magnetic field are obtained varying the relevant input parameters of the model. Section \ref{sec4} is devoted to the study of the mass-radius relation for MSQS by numerically solving the TOV equations. A comparison with some of the available observational data~\cite{Psaltis} is also presented. Finally, our conclusions are given in section \ref{sec5}.

\section{SQM in the presence of a strong magnetic field}
\label{sec2}

In a previous work~\cite{Felipe:2007vb}, the thermodynamical properties of MSQM were studied within the MIT bag model and including the AMM of quarks. Assuming that quarks freely move inside the bag, the total thermodynamical potential is given by
\begin{equation}
\Omega_{total} = \Omega_{vac} + \Omega\,,
\end{equation}
where $\Omega_{vac}=B_{bag} V$ is the QCD vacuum energy density, $V$ is the volume and
\begin{equation}
\Omega= -T \ln Z=\sum_i (\Omega_{i,V}\,V + \Omega_{i,S}\,S +
\Omega_{i,C}\,C),\qquad i=e,u,d,s\,.
\end{equation}
The quantities $\Omega_{V,S,C}$ are the volume, surface and curvature contributions, respectively. Terms associated to the surface and curvature play an important role in processes related to phase transitions, like bubble nucleation in dense matter. Since for the purpose of studying the thermodynamical properties and EoS of SQM in $\beta$-equilibrium only the bulk quark matter is relevant, these two terms can be safely neglected. The explicit form of the volume contribution is
\begin{equation}
\Omega_{i,V}(T,\mu_i) =-\frac{Te_i d_i B}{(2\pi)^3}\sum_{n}\int dk_3
\ln[1+e^{\beta(\mu_i-\epsilon_i)}],
\end{equation}
where $n$ denotes the Landau level, $\mu_i$ is the chemical potential of $i$-particle; $d_e=1$ and $d_{u,d,s}=3$ are degeneracy factors. The quantity $\epsilon_{i}$ corresponds to the energy of the constituents,
\begin{equation}
    \epsilon_{i}^2= k_3^2 + m_{i}^2 \left( \sqrt{
    \frac{B}{B^c_i}(2n+1-\eta) + 1} - \eta y^{i}B \right)^2,
\end{equation}
where $ B^{c}_i=m_{i}^2/|e_i|$ is the critical magnetic field, $y_i=|Q_i|/m_i$
accounts for the AMM and $\eta=\pm 1$ correspond to the orientations of the  particle magnetic moment, parallel or antiparallel to the magnetic field.

For a degenerate MSQM, the energy density, pressures and number density are given by the expressions~\cite{Felipe:2007vb}
\begin{equation}
\varepsilon=B\sum_i{\mathcal M}^0_{i}\sum_{n}\left ( x_ip^{\pm}_{F,i}+h_{i}^{\pm\,\,2}\ln\frac{x_i+p_{F,i}^{\pm}}{h_{i}^{\pm}}\right ),\label{TQi}
\end{equation}
\begin{equation} \label{Ppar}
  P_{\parallel}=B\sum_i {\mathcal M}^0_{i}\sum_n\left ( x_{i}p_{F,i}^{\pm} - h^{\pm\,\,2}_{i}\ln\frac{x_{i} +
 p^{\pm}_{F,i}}{h^{\pm}_{i}}\right ),\end{equation}
\begin{equation}\label{Pper}
 P_{\perp} = B\sum_i {\mathcal M}^0_{i} \sum_{n}\left (2h_{i}^{\pm}\gamma_i^{\pm} \ln\frac{x_{i} + p_{F,i}^{\pm}}{h_{i}^{\pm}}\right),
\end{equation}
\begin{equation}
 N= \sum_i N^0_{i}\frac{B}{B^c_{i}}\sum_{n} p^{\pm}_{F,i}\,,
\qquad {\mathcal M}^0_{i}= \frac{e_i d_i m_{i}^2}{4\pi ^2},
\qquad N_{i}^0 =  \frac{d_i m_{i}^3}{2\pi^2}\,,\label{TQf}
\end{equation}
where the different expressions for the parallel pressure $P_{\parallel}$ and the transverse pressure $P_{\perp}$ reflect the anisotropy of pressures due to the magnetic field. In equations~(\ref{TQi})-(\ref{TQf}) we have defined the dimensionless quantities
\begin{eqnarray} \label{dimvar}
x_{i}=\mu _{i}/m_{i}, \qquad   h_{i}^{\eta} =
\sqrt{\frac{B}{B^{c}_i}\, (2n + 1-\eta) +1} -\eta y_{i}B\,,\nonumber\\
p_{F,i}^{\eta} = \sqrt{x_{i}^2-h_{i}^{\eta}\;^2},\qquad
\gamma^{\eta}_i=\frac{B\,(2n+1-\eta)}{2B^c_i\sqrt{(2n+1-\eta)B/B^c_i+1}}-\eta y_i B\,,
\end{eqnarray}
where $x_i$ is the dimensionless chemical potential, $p_{F,i}$ corresponds to the modified Fermi momentum due to the magnetic field and $h_i$ corresponds to the magnetic mass. The sum over the Landau levels $n$ is up to $n_{max}^{i}$ given by the expression $n_{max}^i = I\left[\left((x_{i} +  \eta y_iB)^2 -1\right)\,B^{c}_i/(2B)\right]$, where $I[z]$ denotes the integer part of $z$.

Let us remark that the quantities defined in (\ref{TQi})-(\ref{dimvar}) contain the contribution of Landau diamagnetism (given by the quantization of Landau levels) as well as the Pauli paramagnetism (due to the presence of AMM for quarks)~\cite{Felipe:2007vb}. Since the inclusion of AMM will not change our main conclusions (it just places a more restrictive upper bound on the magnetic field), in what follows we shall neglect their contribution and consider only the effect of Landau diamagnetism. This can be easily done by taking $y_i=0$ in all the expressions.

\section{Stability window for magnetized SQM}
\label{sec3}

In this section we study the stability window of SQM in a strong magnetic field as a function of the input parameters of the model: the baryon density $n_B$, the magnetic field $B$, the bag parameter $B_{bag}$ and the quark mass $m_s$. In the context of the MIT bag model, and in the absence of a magnetic field, the stability condition for SQM requires the study of the equations
\begin{equation} \label{stabeqs}
P = \sum_i P_i - B_{\rm bag} \label{P_bag}\,,\quad
\varepsilon=\sum_i \varepsilon_i + B_{\rm bag}\,,
\end{equation}
under the condition $P=0$, where $P$ and $\varepsilon$ are the total pressure and energy, respectively. For SQM with massless quarks one obtains the well-known EoS $P=(\varepsilon-4B_{bag})/3$, which at vanishing pressure then leads to the stability condition $\varepsilon=4B_{bag}$.

In the presence of a strong  magnetic field, the anisotropy in the pressures implies $P_{\perp}<P_{\parallel}$. Thus, the stability condition for strong fields changes from $P=0$ to $P_{\perp}=0$ or, equivalently,
\begin{equation} \label{stabpper}
P_{\perp} =\sum_{i}P_{\perp,i}- B_{Bag}=0\,.
\end{equation}
As it turns out from our numerical analysis (see figures below), in this case the total energy $\varepsilon$ given in (\ref{stabeqs}) is always lower than $4B_{bag}$ and, therefore, MSQM is more stable than non-magnetized SQM.


\begin{figure}[t]
\begin{center}
\includegraphics[width=0.6\textwidth]{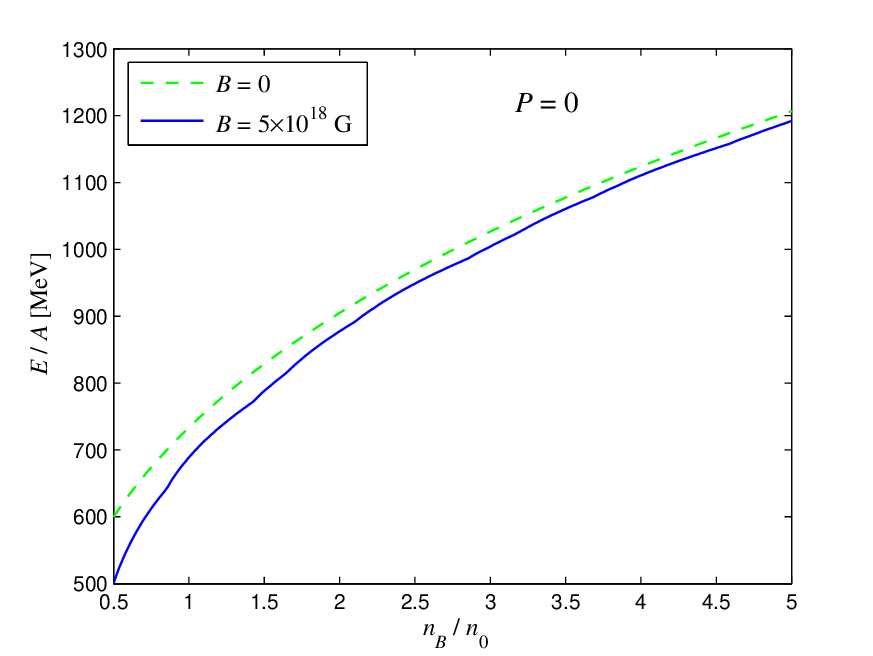}
\end{center}
\caption{Energy per baryon as a function of the baryon density for $B=0$ and $B=5\times 10^{18}$~G, assuming the stability condition of a vanishing pressure.}\label{Energy}
\end{figure}

\begin{figure}[ht]
\begin{center}
\includegraphics[width=0.6\textwidth]{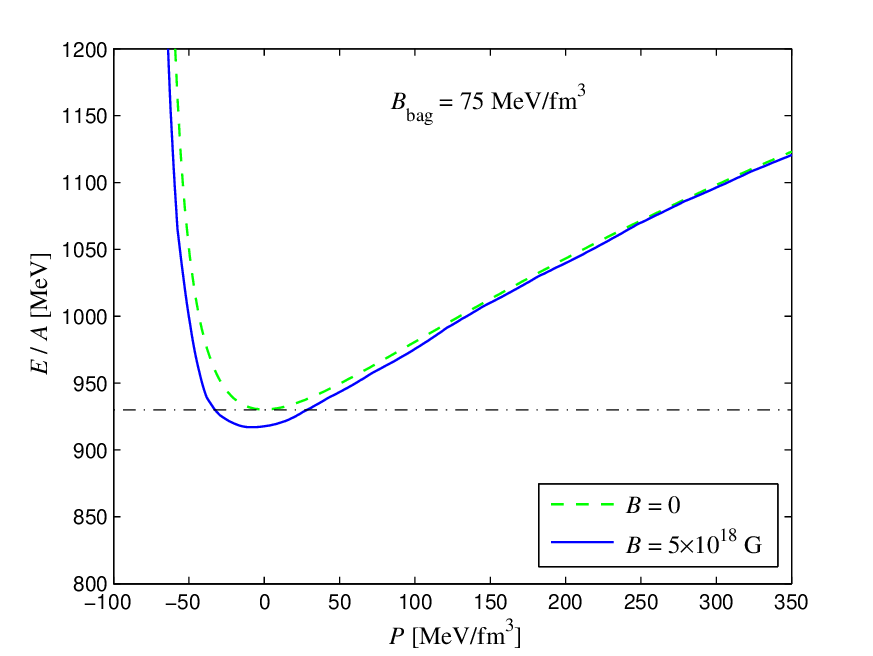}
\end{center}
\caption{Energy per baryon versus pressure for $B=0$ and $B=5\times 10^{18}$~G. We take $B_{\rm bag}=75$~MeV/fm$^3$. The horizontal dot-dashed line corresponds to  $E/A\,(^{56}{\rm Fe}) \simeq 930$~MeV.}\label{EP}
\end{figure}

\begin{figure}[h]
\begin{center}
\includegraphics[width=0.6\textwidth]{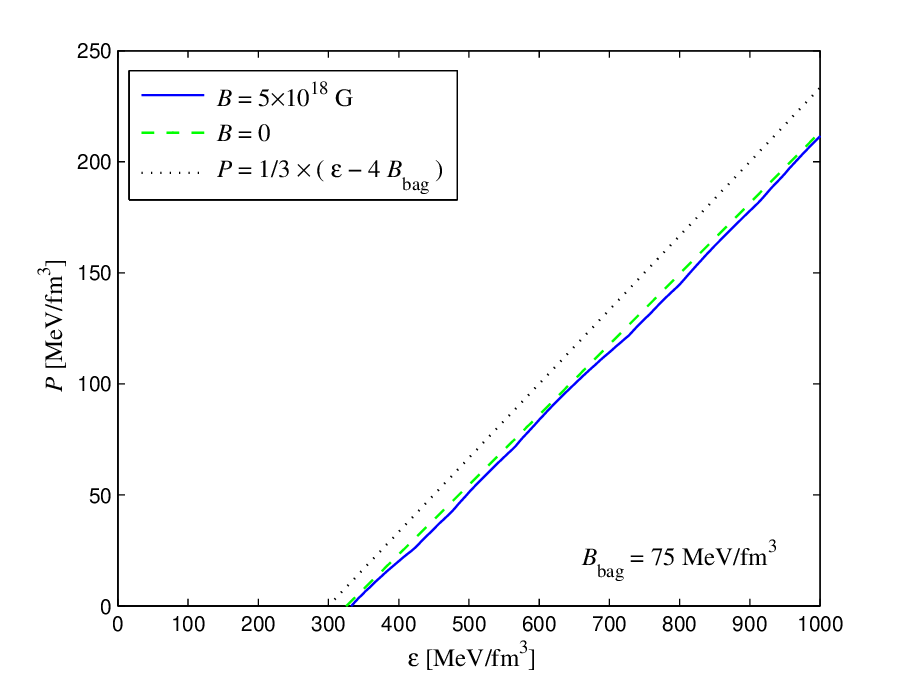}
\end{center}
\caption{EoS: Pressure versus energy for $B=0$ and $B=5 \times 10^{18}$~G. For comparison, the SQM EoS with massless quarks, i.e. $P=(\varepsilon-4B_{\rm bag})/3$, is also shown.}\label{EoS}
\end{figure}

In order to obtain the EoS of MSQM in the stellar scenario under consideration (i.e. the inner core of neutron stars or quark stars) we should also take into account a set of equilibrium conditions: $\beta$-equilibrium, baryon number conservation and electric charge neutrality. Lepton number is not conserved because neutrinos are assumed to enter and leave the system freely ($\mu_\nu=0$). The equilibrium conditions are then determined by the following relations:
\begin{eqnarray}
\mu_u+\mu_e=\mu_d\,,\quad \mu_d=\mu_s\,,\nonumber\\
2N_u-N_d-N_s-3N_e=0\,,\label{beta}\\
N_u+N_d+N_s=3\, n_{B}\,.\nonumber
\end{eqnarray}
The system of equations (\ref{stabpper})-(\ref{beta}) involves all the parameters of the MSQM model. The system is completely determined: there are five equations to find the quark and electron chemical potentials and the bag parameter $B_{bag}$.

In figure~\ref{Energy} we present a comparison of the energy per baryon $E/A$ (or, equivalently, $\varepsilon/n_{B}$) versus the number density $n_B/n_0$ ($n_0 \simeq 0.16$~fm$^{-3}$) in the absence of a magnetic field and for a magnetic field value of $5\times 10^{18}$~G, both at vanishing pressure. We assume $m_u = m_d = 5$~MeV and $m_s =150$~MeV. As can be seen from the figure, $E/A$ is always lower in the presence of a magnetic field. Figure \ref{EP} shows the behavior of the energy per baryon $E/A$ with the pressure $P$, for $B=0$ and $B=5\times 10^{18}$~G. The bag parameter has been chosen to be $B_{\rm bag}=75$~MeV/fm$^3$. One can notice that the point of zero pressure for MSQM is reached for an energy density value lower than that of SQM. Consequently, matter is indeed more stable and more bound when a magnetic field is present. At the zero-pressure point the corresponding baryon density is $n_{B} \simeq 2.18\, n_0$ for $B=0$ and $n_{B} \simeq 2.08\, n_0$ for $B=5\times 10^{18}$~G.

Aiming at obtaining the MSQS observables (e.g. the star mass and radius), next we determine the EoS of MSQM. Since in the present case we do not have a simple algebraic relation between the pressure $P$ and the energy density $\varepsilon$, such a relation must be found numerically. In figure~\ref{EoS} we show the EoS for SQM (i.e. when $B=0$) and for MSQM for a magnetic field around $5\times 10^{18}$~G. For comparison, we also include the EoS of SQM neglecting all the quark masses: $P=(\varepsilon-4B_{\rm bag})/3$. Although the changes in the EoS are not too significant, they can alter the macroscopic observables of stars formed by MSQM, as we shall see in the next section.

\begin{figure}[th]
\begin{center}
\includegraphics[width=0.6\textwidth]{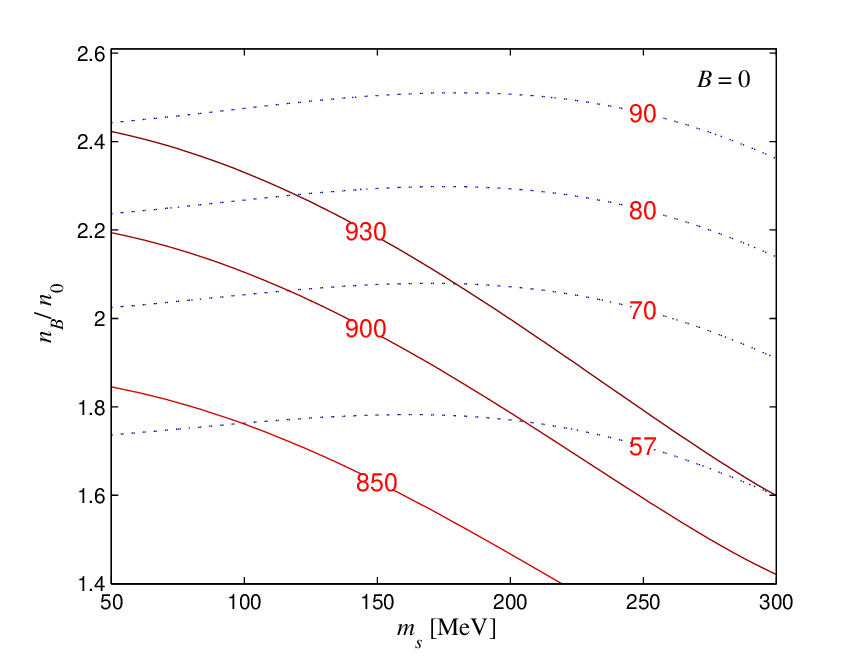}
\end{center}
\caption{The stability windows of SQM in the ($m_s,n_B$)-plane. The contours of constant $B_{\rm bag}$ (dashed lines) and $E/A$ (solid lines) are shown.}\label{cBag}
\end{figure}

\begin{figure}[h]
\begin{center}
\includegraphics[width=0.6\textwidth]{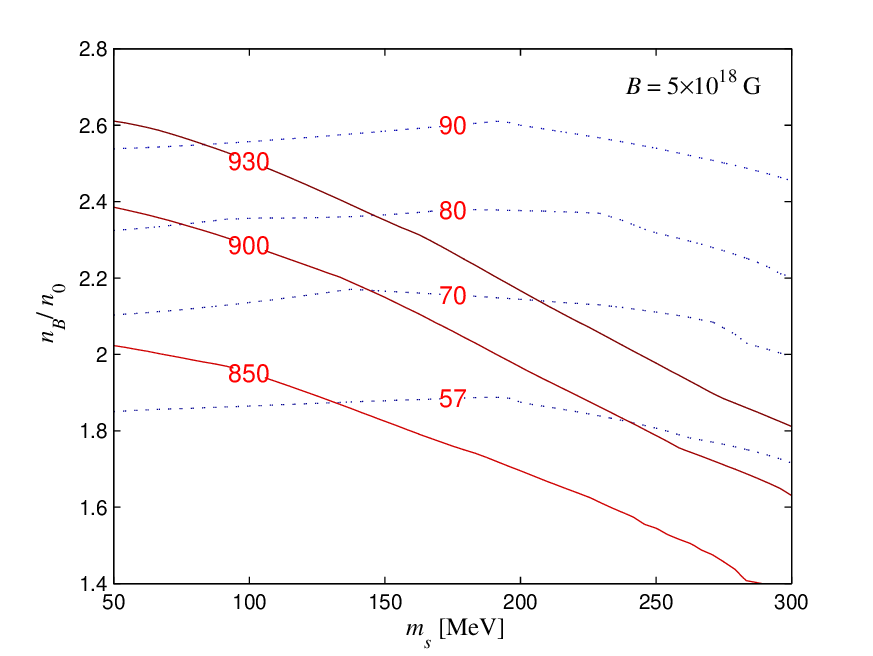}
\end{center}
\caption{The stability windows of MSQM in the ($m_s,n_B$)-plane for $B=5\times 10^{18}$~G. The contours of constant $B_{\rm bag}$ (dashed lines) and $E/A$ (solid lines) are shown.}\label{cE}
\end{figure}

Let us now study the stability window for strange quark matter and, in particular, the regions in the ($m_s,n_B$)-plane where SQM is stable. In order to investigate how the magnetic field affects this window, we fix the magnetic field to $B=5 \times 10^{18}$~G and study the contours of $B_{\rm bag}$ and $E/A$. The results are then compared with the ones obtained for SQM at $B=0$. In figure~\ref{cBag} we present the contours of constant bag parameter and energy per baryon for SQM. The corresponding contours for MSQM are depicted in figure~\ref{cE}.

\begin{table} 
\caption{Bag parameter and baryon density in the presence of a strong magnetic field. We assume $m_s=150$~MeV and $E/A=930$~MeV.}
\label{table1}
\begin{indented}
\item[] \begin{tabular}{ccc}
\br
$B$ (G)& $B_{\rm bag}$ (MeV/fm$^3$) & $n_B/n_0$ \\
\mr
0 & 75  & 2.18 \\
$5\times 10^{18}$ & 80 & 2.37 \\
\br
\end{tabular}
\end{indented}
\end{table}

As can be seen from the figures, the magnetic field tends to shift the stability window of SQM towards higher values of the baryon density (cf. table~\ref{table1}). Indeed, for a given strange quark mass, the constant lines of $B_{\rm bag}$ and $E/A$ in figure~\ref{cBag} are displaced to a higher value of $n_B/n_0$, when compared with the corresponding one of figure~\ref{cE}. While for SQM the allowed range for the baryon density is $1.8 \lesssim n_B/n_0 \lesssim 2.4$ for $50 \leq m_s \leq 300$~MeV and $B_{\rm bag} \gtrsim 57$~MeV/fm$^3$, MSQM allows densities in the range $1.85 \lesssim n_B/n_0 \lesssim 2.6$ for $50 \leq m_s \leq 240$~MeV. The approximate lower bound of 57~MeV/fm$^3$ for the bag parameter is the limit imposed by requiring instability of two-flavor quark matter (deconfined gas of $u$ and $d$ quarks)~\cite{Farhi:1984qu,buballa}. We also notice that the allowed interval for the bag parameter is affected by the magnetic field. Below the energy contour of 930 MeV, our EoS for MSQM corresponds to an $E/A$ at $P=0$ lower than that of $^{56}$Fe and it can therefore be considered absolutely stable.

\section{SQS mass-radius relation}
\label{sec4}

In this section we study the equilibrium configuration of magnetized strange quark stars described by the EoS of MSQM obtained in the previous section. As is well known, the most important macroscopic parameters of a star are its radius $R$ and its gravitational mass $M$. The surface redshift $z_s$ is also an important parameter which is related to the star mass and radius through the equation 
\begin{equation}\label{redshift}
z_s=\frac{1}{\sqrt{1-\frac{2 G M}{R c^2}}}-1,
\end{equation}
where $G$ is the gravitational constant. The interest in this parameter is that it is observable.

Configurations of spherical symmetric non-rotating compact stars are obtained by  the numerical integration of the TOV equations~\cite{Baym},
\begin{eqnarray}
\frac{dM}{dr}=4\pi G r^2\varepsilon (r)\,,\nonumber\\
\frac{dP}{dr}=
-G\frac{\left(\varepsilon (r) +P (r)\right)\left( M(r) + 4\pi P(r) r^3 \right)}{r^2-2 r M (r)}\,,
\label{TOV}
\end{eqnarray}
supplemented with the EoS, where $P(r)$ is the pressure and $\varepsilon(r)$ is the energy density. The radius $R$ and the corresponding mass $M$ of the star are determined by the value of $r$ for which the pressure vanishes, $P(R)=0$. The EoS fixes the central pressure, $P(0)= P_c$, which together with the condition $M(0)=0$, completely determine the system of equations (\ref{TOV}). In this way, varying continuously the central pressure, one obtains a mass-radius relation $M(R)$, which relates masses and radii for a given EoS. The stable branches of these curves must satisfy the condition $dM/dP_c > 0$. Other solutions are unstable and collapse.

\begin{figure}[t]
\begin{center}
\includegraphics[width=0.6\textwidth]{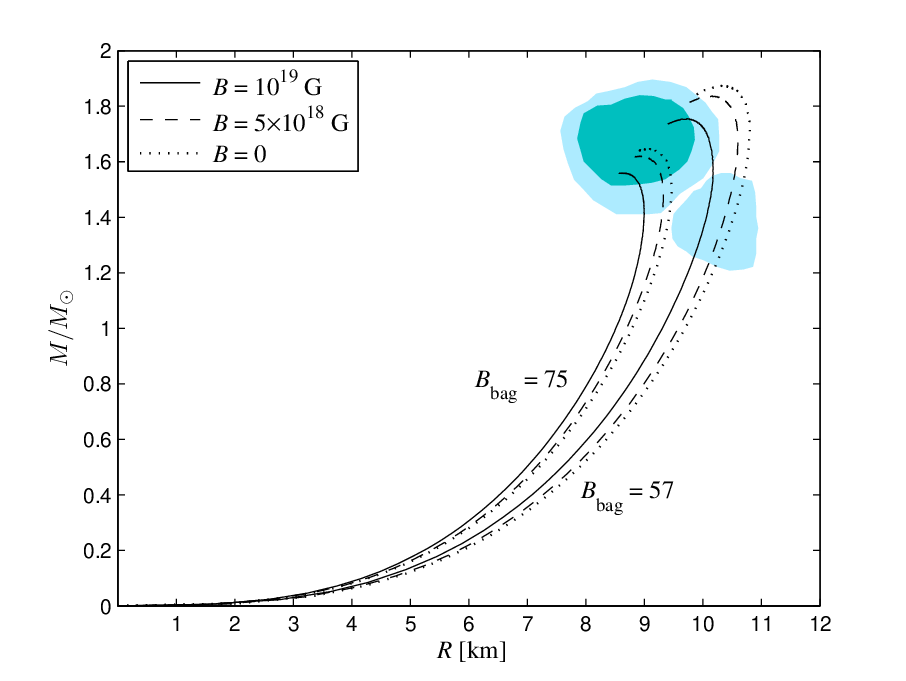}
\end{center}
\caption{Possible $M-R$ configurations obtained for MSQS in the present model. The curves are shown for different values of the magnetic field $B$ and bag parameter $B_{\rm bag}$ (in units of MeV/fm$^3$). The shaded (blue) areas correspond to $1\sigma$ and $2\sigma$ confidence contours for the mass and radius of the X-ray binary EXO 1745-248~\cite{Psaltis}.}\label{tov}
\end{figure}

Let us now study the static stellar configurations for MSQS. We remark that, despite the anisotropy of pressures in the presence of the magnetic field, this effect is not significant for $B \lesssim 10^{19}$~G. Therefore, in our astrophysical context, even if the EoS is determined by the condition (\ref{stabpper}), TOV equations can be solved using a spherical metric, which then leads to equations (\ref{TOV}). 

Starting from the numerical EoS of MSQM, we have solved the TOV equations to obtain
the stable mass-radius configurations of MSQS. As mentioned before, the magnetic field enhances the stability of SQM and, thus, it allows the appearance of 
stable stars with masses and radii smaller than for non-magnetized SQM.

In figure~\ref{tov} we plot the stable configurations of SQM without a magnetic field, as well as of MSQM, for two different values of the magnetic field, $B=5\times 10^{18}$~G and $10^{19}$~G. The curves are presented for two values of the bag parameter, $B_{\rm bag}=57$~MeV/fm$^3$ and 75~MeV/fm$^3$. Our main results are summarized in tables \ref{table2} and \ref{table3}, where the maximum mass $M_{\rm max}$ and maximum radius $R_{\rm max}$ of MSQS are given. 

For the sake of comparison with observational data, we have also depicted in figure~\ref{tov} the $1\sigma$ and $2\sigma$ contours for the mass and radius of the low mass X-ray binary EXO 1745-248, based on the spectroscopic data during thermonuclear bursts combined with a distance measurement to the globular cluster~\cite{Psaltis}.
We observe that the curves predicted by the MSQM EoS are in good agreement with the confidence contours.

\Table{\label{table2}Maximum mass and the corresponding radius of a MSQS for different values of the magnetic field and bag parameter.}
\br
&\centre{2}{$B_{\rm bag}=57$}&\centre{2}{$B_{\rm bag}=75$}\\
\ns
&\crule{4}\\
$B$~(G)&$M_{\rm max}/M_{\odot}$&$R$ (km)&$M_{\rm max}/M_{\odot}$&$R$ (km)\\
\mr
0 &1.87&10.35 &1.65&9.07\\
$5\times 10^{18}$&1.84 &10.16 &1.62&8.91\\
$10^{19}$&1.76 &\;\,9.73 &1.56 & 8.62\\
\br
\end{tabular}
\end{indented}
\end{table}

\Table{\label{table3}Maximum radius and the corresponding mass of a MSQS for different values of the magnetic field and bag parameter.}
\br
&\centre{2}{$B_{\rm bag}=57$}&\centre{2}{$B_{\rm bag}=75$}\\
\ns
&\crule{4}\\
$B$~(G)&$R_{\rm max}$\,(km)& $M/M_{\odot}$&$R_{\rm max}$\,(km)&$M/M_{\odot}$\\
\mr
0 & 10.80& 1.73 &9.47 &1.49 \\
$5\times 10^{18}$& 10.60 &1.69  &9.32 & 1.50\\
$10^{19}$&10.17  &1.58 & 9.00 & 1.41 \\
\br
\end{tabular}
\end{indented}
\end{table}

\section{Conclusions}
\label{sec5}

In the present work, we have investigated the stability of magnetized strange quark matter within the phenomenological MIT bag model. We have studied the stability windows of MSQM taking into account the variation of the strange quark mass, the baryon density, the magnetic field and the bag parameter. We found that MSQM is indeed more stable than non-magnetized SQM. While for SQM the stability range for the baryon density is $1.8 \lesssim n_B/n_0 \lesssim 2.4$ for $50 \leq m_s \leq 300$~MeV, MSQM allows densities in the range $1.85 \lesssim n_B/n_0 \lesssim 2.6$ for $50 \leq m_s \leq 240$~MeV and a magnetic field value of $5\times 10^{18}$~G. Moreover, the allowed range for the bag parameter is $57 \lesssim B_{\rm bag} \lesssim 90$~MeV/fm$^3$.

At vanishing pressure and finite density, the derived EoS for MSQM exhibits a minimum of the energy per baryon lower than that of $^{56}$Fe. This gives the possibility of existence of stable configurations for magnetized strange quark stars. In this simple framework for MSQM, we have then derived the mass-radius relation for such compact stars and compared the theoretically predicted relations with some of the recent observational data. We have concluded that in the presence of a strong magnetic field, stable strange stars with smaller masses and radii could exist.

\ack{
The authors acknowledge the fruitful discussion with J. Horvath and his important suggestions and comments. A.P.M. thanks CFTP-IST~(Lisbon,Portugal) for their hospitality and financial support. This work was partially supported by \emph{Funda\c c\~ ao para a Ci\^ encia e a  Tecnologia} (FCT, Portugal) through the projects POCI/FP/81919/2007, and CFTP-FCT UNIT 777,  which are partially funded through POCTI (FEDER). The work of R.G.F. has been partially supported by the Marie Curie Research Training Network MRTN-CT-2006-035505. The work of A.P.M. has been supported by \emph{Ministerio de Ciencia, Tecnolog\'{\i}a y Medio Ambiente} under the grant CB0407 and the ICTP Office of External Activities through NET-35.  A.P.M. also acknowledges TWAS-UNESCO for financial support at CBPF-Brazil.\\}

\end{document}